\documentclass{emulateapj}

\shorttitle{The effect of a stellar magnetic variation on the jet velocity}
\shortauthors{De Colle et al.}

\begin{document}

\title{The effect of a stellar magnetic variation on the jet velocity}

\author{Fabio De Colle\altaffilmark{1}, Jos\'e Gracia\altaffilmark{1}}
\affil{Dublin Institute for Advanced Studies (DIAS), 31 Fitzwilliam Place, Dublin 4, Ireland}

\and
\author{Gareth Murphy\altaffilmark{2}}
\affil{Laboratoire d'Astrophysique de Grenoble, CNRS, Universit\'e Joseph Fourier, B.P. 53, F-38041 Grenoble, France}

\begin{abstract}

Stellar jets are normally constituted by chains of knots with some 
periodicity in their spatial distribution, corresponding to 
a variability of order of several years in the ejection from the protostar/disk 
system.
A widely accepted theory for the
presence of knots is related to the generation of internal working
surfaces due to variations in the jet ejection velocity.
In this paper we study the effect of variations in the inner 
disk-wind radius on the jet ejection velocity.
We show that a small variation in the inner disk-wind radius
produce a variation in the jet velocity large enough
to generate the observed knots.
We also show that the variation in the inner radius may be related
to a variation of the stellar magnetic field.

\end{abstract}



\section{Introduction}

Stellar jets are observed in
the form of chains of emitting nebulae called Herbig-Haro (HH) objects.
HH objects are generally thought to be shock-heated density
condensations traveling along the outflows formed by star-disk
systems during the process of star formation \citep[e.g.][]{rei01}.

Since their discovery several scenarios have been suggested for the 
origin of \emph{steady} outflows from young stellar objects.
In the stellar wind model, material is accelerated by thermal pressure 
gradients \citep[e.g.][]{can80}. 
Magnetohydrodynamics (MHD) models rely on the magnetocentrifugal launching
mechanism \citep{bla82}.
For the X-wind scenario the
jet is magnetically driven from the so-called ``X-annulus'' where the young star's
magnetosphere interacts with the disk \citep{shu00}.
In the disk wind scenario the jet is launched from an extended 
region of the disk surface \citep[e.g.][]{fer97}.
The analytical models mentioned above have focused mainly 
on the steady-state aspect of the ejection phenomena. 

\emph{Unsteady} periodic ejections  
with timescales of the order of several rotation periods of the inner disk radius
have been obtained by numerical simulations \citep[e.g.][]{ouy97,goo99,mat02}.
Observations indicate a significantly longer timescale is associated with the
appearance of knots in stellar outflows. 
Nearly all observed jets present small scale knots
 up to 0.1 pc from the central source, with a spacing between the knots 
 corresponding to a timescale of $\approx$ 1-20 yr 
(e.g. HH30 - \citealt{bur96}; HH111 - \citealt{rei92}; RW-Aur - 
\citealt{lop03}). More 
fragmented knots are observed on a typical timescale of $\sim 10^{2-3}$ yr
at larger distances from the source.
While the long term variation of jets can be 
explained by variations in the accretion rates (e.g. during FU-Orionis phases),
the possible origin of small scale knots is still unclear.

A first possibility, suggested by similarities with extra-galactic jets,
is that the knots may be formed by hydrodynamics Kelvin-Helmholtz instabilities \citep{mic98},
MHD Kelvin-Helmholtz reflective pinch mode instabilities \citep[]{cer99},
or by current-driven instabilities \citep[]{fra00}.
Numerical simulations by these authors have shown that the shocks generated
by plasma instabilities are weaker than those seen in observations once 
radiative cooling is taken into account.
Moreover, some jets \citep[e.g. HH212 -][]{zin98} show a remarkable symmetry
on both sides of the central star-disk system. This may be difficult to explain
assuming that the instabilities are triggered by small scale perturbations.
Additionally, and more importantly, optical images in many cases show 
that the compact knots have a bow-shock structure (e.g. HH111 - \citealt{rei92}) 
that is difficult to obtain supposing that the knots are formed
by instabilities.

A widely accepted theory for the presence of knots in
stellar jets is related to the generation of internal working surfaces due 
to supersonic variations of the ejection velocity at the base of the 
jet \citep{rag90}.
Indeed, numerical simulations, using a velocity variations
of about 10-20\% of the average velocity,
have been able to reproduce  
the morphology and the emission property of the knots in HH objects 
\citep[e.g.][]{esq07}.

In this paper 
we show that velocity variations which lead to the creation of knots similar to 
the observed ones may be generated by variations of the inner disk-wind 
radius. Furthermore, we suggest that these variations may be related
to changes in the stellar magnetic field.

This paper is organized as follows.
In Section \ref{model} we determine the effect of a
variation in the inner disk radius on the jet velocity.
In Section \ref{variable}
we suggest that the variation in the inner radius
could be connected to a
change of the stellar magnetic field.
In Section \ref{discussion} we discuss the approximation used and 
the limitations of our model. 
Conclusions are given in Section \ref{conclusion}.


\section{The effect of a variable inner ejection radius on the jet velocity}
\label{model}

We first discuss some
asymptotic properties of a single fieldline anchored to 
an accretion disk as a function of the conserved invariants
and the fieldline's footpoint radius. 
Then we extend the discussion to an ensemble of fieldlines
anchored at a range of footpoint radii and derive mean values for
the asymptotic velocity of a disk-wind extending from an inner to
an outer finite radius. Lastly, we discuss the consequences of a
time-dependent inner disk-wind radius on the average velocity of 
the outflow.

\subsection{Asymptotics along a single fieldline}

In steady state, axisymmetric, ideal MHD a number of quantities are
conserved along a fieldline, i.~e. a surface of constant magnetic flux
$\Psi$. Among these invariants are the
mass-to-magnetic flux ratio $k(\Psi)$, the total angular momentum
$L(\Psi)$ and the corotation frequency or angular velocity
$\Omega(\Psi)$ (e.g. \citealt{bla82}; \citealt{pel92}). 
These invariants are defined as
\begin{equation} \label{eq:streamfunc}
  k(\Psi) = 4\pi\rho \, \frac{V_{\rm p}}{B_{\rm p}} \,,
\end{equation}
\begin{equation} \label{eq:angmom}
  L(\Psi) = R \left( V_\phi - \frac{B_\phi}{k(\Psi)}\right) \,,
\end{equation}
\begin{equation} \label{eq:Omega} 
  \Omega(\Psi) = \frac{1}{R} \left(V_\phi - \frac{k(\Psi)}{4\pi\rho} B_\phi\right) \,,
\end{equation}
where $\rho$ is the mass density, $V_{\rm p}$, $V_\phi$, $B_{\rm p}$, $B_\phi$
are the poloidal and toroidal components of the velocity and magnetic field,
and $R$ is the distance from the star.
The invariants $k(\Psi)$, $\Omega(\Psi)$, $L(\Psi)$ and the footpoint $R_0(\Psi)$ 
are assumed to be known or given functions.

$L(\Psi)$ includes contributions from the twisted magnetic
field.
The relation
\begin{equation} \label{eq:Ralfven}
  L(\Psi) = R^2_{\rm A}(\Psi) \, \Omega(\Psi) 
\end{equation}
holds, where $R_{\rm A}(\Psi)$ is the Alfv\'en radius. 
Eq. (\ref{eq:Ralfven}) 
determines $R_{\rm A}(\Psi)$ and subsequently
$\lambda(\Psi) = R_{\rm A}(\Psi)/R_0(\Psi)$ (the magnetic lever
arm) completely. 
Here, and in the
following, all quantities with an explicitly indicated functional
dependency on $\Psi$, e.g. $L(\Psi)$, are taken to depend on the magnetic
flux $\Psi$ only.  Similarly, all quantities with a subscript $_0$ are
taken to be defined at the base of the fieldline, i.~e. the equatorial
plane. The corotation frequency $\Omega(\Psi)$ can be easily
evaluated at the equator -- where the poloidal velocity component
$V_{\rm p}$ vanishes -- and equals the orbital
frequency at the equatorial plane $\Omega_0$:
\begin{equation} \label{eq:Omega_eq}
  \Omega(\Psi) = \Omega_0(\Psi) \,.
\end{equation}

Another conserved quantity is given by the total energy $E(\Psi)$
per unit mass, with
contributions from kinetic, thermal, gravitational energy $\Phi$ and the
energy of the electromagnetic field (Poynting flux):
\begin{equation} \label{eq:totenergy}
  E(\Psi) = \frac{1}{2} V^2 + \frac{\gamma}{\gamma-1} \frac{P}{\rho} 
         + \Phi - \frac{\Omega(\Psi) R B_\phi}{k(\Psi)} \,,
\end{equation}
where $P$ is the thermal pressure and $\gamma$ is the ratio of specific heats.
The asymptotic jet velocity along a fieldline $\Psi$ anchored at the
footpoint radius $R_0(\Psi)$ can be estimated from the 
Bernoulli equation for large distances and negligible enthalpy as \citep{mic69}:
\begin{equation} \label{eq:michel}
  v_{\infty}(\Psi)
  \sim \sqrt{2} \, \Omega(\Psi) \, R_0(\Psi) \, \lambda(\Psi) \,.
\end{equation}

The asymptotic jet velocity varies from fieldline to
fieldline. However, we are ultimately interested in a typical
value representative of the jet as a whole. Therefore, we will
weigh the asymptotic velocity with the mass-flux carried by the
fieldline. 

The asymptotic mass-flux along a fieldline per unit magnetic flux 
(for one hemisphere) can be estimated by rewriting eq. (\ref{eq:streamfunc}) as
\begin{equation} \label{eq:massflux}
   \frac{\partial\dot{M}}{\partial\Psi} 
   = \frac{k(\Psi)}{2} \,,
\end{equation}
where $\dot{M}$ is the wind mass-loss rate.

\subsection{Asymptotic mean jet velocity}

The jet is assumed to originate in a disk-wind. Mass loading onto the disk-wind
shall be efficient only for magnetic flux surfaces in the range $\Psi_{\rm in}
\le \Psi \le \Psi_{\rm ex}$, where the flux surfaces $\Psi_{\rm in}$ and $\Psi_{\rm ex}$
are anchored at $R_{\rm in} = R_0(\Psi_{\rm in})$ and $R_{\rm ex} = R_0(\Psi_{\rm ex})$,
respectively.
The mean or typical velocity of the jet given by the
superposition of the asymptotic velocities along the fieldlines
anchored between $R_{\rm in}$ and $R_{\rm ex}$ can be determined
weighting the jet velocity by the mass-flux carried
along every fieldline:
\begin{equation}
 \langle v \rangle = \frac{\int_{R_{\rm in}}^{R_{\rm ex}} v_\infty d\dot{M}}
               {\int_{R_{\rm in}}^{R_{\rm ex}} d\dot{M}} \,.
 \label{eq:var}
\end{equation}

So far, we have assumed steady state and axisymmetry. 
In the following we will add two
assumptions, namely keplerian rotation and self-similarity \cite[see e.g.][]{VT98}. 

Assuming keplerian rotation for the plasma in the equatorial plane
fixes the corotation frequency to
\begin{equation} \label{eq:Omega_final}
  \Omega(\Psi) = \Omega_0(R_0(\Psi)) = \sqrt{\frac{GM}{R_0^3(\Psi)}} \,.
\end{equation}
The magnetic flux is a power-law in $\Omega$, i.~e. $\Psi \sim
\Omega^{-\alpha}$. Self-similarity fixes the power-law
index to $\alpha = 1/2$ \citep{bla82}.
In terms of the footpoint radius $R_0$, $\Psi$ is given by
\begin{equation}
  \Psi = \Psi_{\rm ex} \, (R_0/R_{\rm ex})^{3/4} \,.
  \label{eq:psi}
\end{equation}

\begin{figure}
 \centering
 \includegraphics[width=84mm]{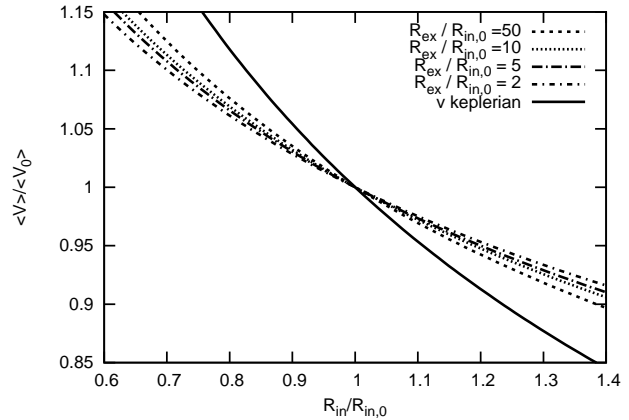}
  \caption{Average outflow velocity as a function of $R_{\rm in}/R_{\rm in,0}$.
           The average outflow velocity is normalized to $\langle v_0 \rangle = \langle v \rangle_{R=R_{\rm in,0}}$.}
  \label{fig2}
\end{figure}

Self-similarity also fixes the mass-to-magnetic flux function $k(\Psi)$ to
\begin{equation}
  k(\Psi) = k_{\rm ex} \, (R_0/R_{\rm ex})^{-3/4} \,,
  \label{eq:k}
\end{equation}
and renders the lever arm independent of the magnetic
flux surface (i.~e. $\lambda(\Psi) = \lambda$).
Keplerian rotation
and constant $\lambda$ is sufficient to completely determine the
asymptotic velocity along a fieldline as
\begin{equation} \label{eq:vjet}
  v_{\infty}(\Psi) 
  = \lambda \sqrt{\frac{2GM}{R_0(\Psi)}} \,,
\end{equation}
while
the wind mass-loss rate is determined by eq. \ref{eq:massflux} and yields
\begin{equation}
 d\dot{M}=\frac{3}{8} k_{\rm ex} \Psi_{\rm ex} \frac{d R_0}{R_0} \,.
 \label{eq:mdot}
\end{equation}

Finally, defining $\chi = R_{\rm ex}/R_{\rm in}$, the average velocity (eq. \ref{eq:var}) 
can be integrated, giving:
\begin{equation}\label{eq:vmean}
  \langle v \rangle =  2 \; 
  \frac{\chi^{1/2}-1}{\ln\chi} \; v_\infty(R_{\rm ex}) \,.
\end{equation}
where $R_{\rm ex}$ is assumed fixed.
The velocity variation as function of the inner radius,
for different values of $\chi$, is shown in Fig. \ref{fig2}.
 As this Figure clearly shows, an increase in the inner radius causes a 
decrease in the average velocity.
Moreover, the velocity variation is nearly independent of the ratio 
$\chi_0=R_{\rm ex}/R_{\rm in,0}$ (where $R_{\rm in,0}$ is a position 
of the inner radius chosen as a reference).
To produce a variation of 20\% in velocity it is necessary to have a variation
of order of 30\% in the inner radius.
Fig. \ref{fig2} shows also the difference between the velocity of the disk-wind 
at the inner radius (keplerian, see eq. \ref{eq:vjet}) and the 
average velocity (defined by eq. \ref{eq:var}).

\begin{figure}
 \centering
 \includegraphics[width=64mm]{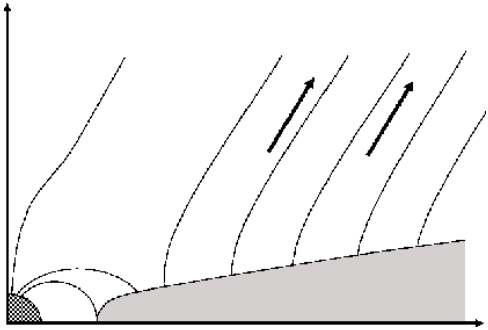}
  \caption{Schematic representation of the 
          studied model, where an extended disk-wind is 
          ejected by an accretion disk with a time-varying
          stellar magnetic field.}
  \label{fig1}
\end{figure}


\section{The effect of a magnetic field cycle on the inner disk-wind radius}
\label{variable}

So far, we have dealt exclusively with the magnetic field of the
disk-wind. These magnetic flux surfaces carrying the disk-wind are
necessarily open to infinity. A non-open fieldline cannot contribute
to the global outflow. In the following we consider the superposition
of a dipolar magnetosphere carried by the central star, contributing
magnetic flux $\Psi_{\rm DP}$ and the previously discussed magnetic field
$\Psi$ which threads the disk and carries the outflow. Then the total
magnetic flux is  given by $\Psi + \Psi_{\rm DP}$.

This configuration leads to three
different kinds of magnetic field lines as is illustrated in Fig.
\ref{fig1}. Firstly, fieldlines anchored in the polar region of
the stellar surface. While these fieldlines are open to infinity, they are
assumed to carry only negligible mass-flux and do not contribute
significantly to the outflow.
Secondly, closed dipolar fieldlines anchored at lower latitudes
of the stellar surface do not contribute to the outflow in our
model. Lastly, a global open magnetic field threading the disk, which
carries the disk-wind.

The transition from the closed dipolar fieldlines to the open
disk-wind, i.~e. the location of the innermost open flux surface $\Psi_{\rm in}$
in the equatorial plane, is given by the location of the saddle point
of the magnetic flux distribution

\begin{equation}
 \left[ \frac{d}{dR_0} \left(\Psi + \Psi_{\rm DP} \right) \right]_{R_0=R_{\rm in}} = 0 \,,
 \label{eq:defRin}
\end{equation}
where the dipolar field on the equatorial plane is given by a
time-varying dipole moment $m(t)$ as
\begin{equation} \label{eq:dipol}
  \Psi_{\rm DP} \propto m(t) \frac{1}{R_0}.
\end{equation}
and $\Psi$ is given by eq. \ref{eq:psi}.
Since both magnetic flux distributions are known, the inner radius of
the disk-wind is given in terms of the time-varying stellar magnetic 
field (defined on the stellar surface) from eq. \ref{eq:defRin} as  
\begin{equation}
  R_{\rm in} \propto |m(t)|^{4/7}.
\end{equation}

For simplicity, the stellar magnetic dipole moment
is assumed to vary in time as
\begin{equation}
  m(t) =  m_0 \left (1 + \delta \, \sin(2 \pi t/\tau_\star) \right)
  \label{eq:bvar}
\end{equation}
where $\tau_\star$ and $\delta$
are the period and the amplitude
of the stellar magnetic field variation, and $m_0$
is the stellar magnetic dipole moment at $t=0$.

The inner radius of the ejecting region depends on the
variation in the magnetic field intensity, and 
changes on timescales of $\tau_\star$.
The average velocity will therefore change periodically following the
magnetic field cyclic variation.

\begin{figure}
 \centering
 \includegraphics[width=84mm]{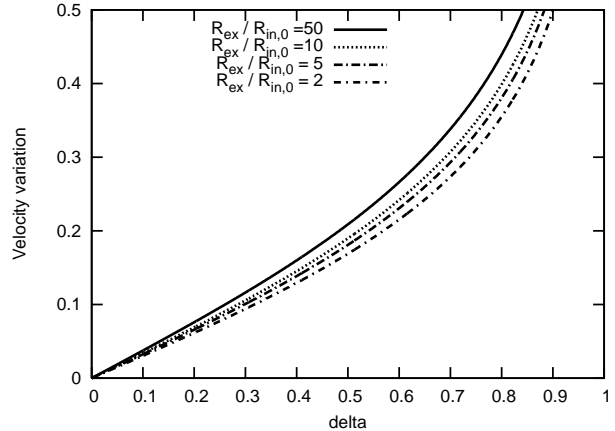}
  \caption{Variation of the velocity (normalized to the average velocity) 
           as a function of $\delta$ (the amplitude of the 
           stellar magnetic field variation) for different values of 
           $\chi_0$.}
  \label{fig3}
\end{figure}

A plot of the velocity variations, that is the difference between
the maximum and minimum velocity, and normalized to 
the average velocity during one cycle of the stellar magnetic field, 
is shown in Fig. \ref{fig3} as function of $\delta$.
The different curves are calculated corresponding to different values 
of $\chi_0$ ($=R_{\rm ex}/R_{\rm in,0}$).
A 50\% variation in the stellar magnetic field
produces a $\sim$ 20\% jet velocity variation,
nearly independent of the value of $\chi_0$.

We have assumed that the only effect  
of a stellar magnetic field variation on the ejected jet is  
a change in $R_\mathrm{in}$.
Actually, some of the open stellar flux will thread the disk.
Therefore, the disk magnetic field (including also the contribution
from the open stellar magnetic flux) will not decrease 
as $\sim R_0^{-5/4}$ and will also change with time
following the change in the stellar dipole.

A change in the stellar magnetic field produces
changes in the mass-flux and velocity.
In fact, the lever arm $\lambda$ is a function of the magnetic field
and of the mass flux (that is also a function of the magnetic field).
\citet{fer97} and \citet{cas00} studied models of self-similar
MHD accretion disks driving jets. These authors showed that 
$\lambda^2 \sim 1 + 1/(2 \xi)$, where $\xi$ is the ejection parameter
(that relates the mass flux to the radius by $\dot{M} \sim r^\xi$).
Additionally, $\xi \sim  0.1 \mu^3$, where $\mu$ is the magnetization parameter
(given by $\mu =B^2/(8 \pi p)$, where $p$ is the thermal pressure).

In this case, eq. \ref{eq:vjet} and \ref{eq:mdot}
may be written as:
\begin{equation}
  v_{\infty}(\Psi) 
  = \lambda(\Psi) \; \sqrt{\frac{2GM}{R_0(\Psi)}} \,, \qquad
 d\dot{M} \propto \xi \; R_0^\xi \; \frac{d R_0}{R_0} 
 \label{eq:mdotd}
\end{equation}
An increase of the magnetic field in the disk
(due to the increased stellar contribution)
produces an increase in the ejection parameter $\xi$ and of
the mass flux, and a drop in the asymptotic velocity.
Therefore, the jet velocity decreases further with respect
to the value calculated by eq. \ref{eq:vmean}.

The average velocity (eq. \ref{eq:var}) is given by:
\begin{equation}
 \langle v \rangle = \frac{\int_{1/\chi}^1 
         v_\infty(R_{\rm ex}) \lambda/\lambda(R_{\rm ex}) \xi x^{\xi-3/2} dx}
            {\int_{1/\chi}^1 \xi x^{\xi-1} dx} \,.
 \label{eq:vard}
\end{equation}
where in the integrals $x=R_0/R_{\rm{ex}}$.
Assuming $\xi$ independent of $x$ (i.~e. self-similarity), 
eq. \ref{eq:vard} leads to:
\begin{equation}
 \langle v \rangle = 2  \;
 \frac{\xi}{1-2\xi} \;
 \frac{\chi^{1/2}-\chi^{\xi}}{\chi^{\xi}-1} \;
 v_\infty(R_{\rm ex})
\,,
 \label{eq:varf}
\end{equation}
where $R_{\rm ex}$ is assumed to be fixed, but
$v_{\infty}(R_{\rm ex})$ varies with the stellar magnetic field
as a function of $\lambda(R_{\rm ex})$
($\sim 1/(2 \xi)^{1/2}$ for $\xi \ll 1$).

For $\xi \rightarrow 0$ (assuming that $\lambda$ is converging
to a large but finite value) eq. \ref{eq:varf} reduces to eq. \ref{eq:vmean}.

The ratio between the external and inner radius $\chi$ is difficult to
determine theoretically, but has been recently constrained
by observations.
In fact, interpreting the observed transverse velocity shifts
in T Tauri microjets as indication of rotation, a range of 
ejecting radii corresponding to $\chi \approx 10$ can be 
inferred \citep[e.g.][]{fer06}.

Therefore, we may assume $\xi \ll 1$, and write the average outflow 
velocity normalized to $\langle v_0 \rangle = \langle v \rangle_{R=R_{\rm in,0}}$ as:
\begin{equation}
 \frac{\langle v \rangle }{\langle v_0 \rangle} \approx 
         \left(\frac{\langle v \rangle}{\langle v_0 \rangle}\right)_{\xi=0} \;
 \sqrt{\frac{\xi_0}{\xi}}
 \label{eq:vv0}
\end{equation}
where the scaling $\langle v \rangle \sim 1/\sqrt{\xi}$ 
comes directly from $v_{\infty} \sim \lambda \sim 1/\sqrt{\xi}$.
From this equation, it becomes clear that
the stellar magnetic field variation necessary to produce 
a 20\% variation in the average velocity
may be much smaller than the 50\% value determined from Fig. \ref{fig3}.
The 50\% value represents an upper limit
to the necessary stellar magnetic field variation.


\section{Discussion}
\label{discussion}

Is this section we discuss in some detail the approximations and 
the limitations of the model.
First, to use the
results from the standard disk-wind theory,
we approximated the evolution of the
system as a sequence of steady-state configurations.

We made two key assumptions 
on the behavior of the stellar magnetic field.
The first assumption of a dipolar stellar magnetic field
at the inner edge of the disk is standard in the
magnetospheric accretion model for young stars \citep{kon91}.
On the stellar surface, magnetic fields as large as several
kG have been observed \citep[e.g.][]{joh07}. The dipole
component ($\sim 100-500$ G on the stellar surface), which decays more slowly 
then the multipole components, dominates at large distances
from the star.

The second hypothesis is that 
a dynamo mechanism operates in protostars \citep{cha06,dob06}
 with a periodic or quasi-periodic variation of the magnetic field with a 
typical timescale of a few to tens of years.
A dynamo mechanism is necessary to explain the observed magnetic field,
because this would be dissipated on scales of order
of $R_\star/\eta \approx 100$ yr, where $R_\star$ is the stellar radius
and $\eta$ is the turbulent magnetic diffusivity \citep[e.g.][]{cha06}.
A dynamo mechanism, if present, is not necessarily cyclic
as needed by our model.
However, recently \citet{sok08} 
proposed models of dynamos in low mass, fully convective stars 
with a cyclic magnetic field.
The photometric variability observed in T Tauri stars 
\citep[e.g.][]{mel05,gra07} may be related to a corresponding
change in the stellar magnetic field (\citealt{arm95}).

Finally, our results depend on the complex
interaction between the stellar magnetic field and the disk.
This has 
been studied by a number of authors both analytically and 
numerically (see, e.g., \citealt{uzd04} and references therein).
It is well understood that the differential rotation 
in the disk produces a winding-up of the magnetic field lines
on a timescale $ \tau = 2 \pi/\left(\Omega_{_\star}-\Omega\right)$
(where $\Omega_{_\star}$ is the stellar rotation frequency)
with a following opening of the stellar magnetic field lines
and a consequent outflow.

The later evolution of the system is unclear.
Once the magnetic field lines are opened, they can stay in that
configuration indefinitely \citep{lov95} allowing a 
disk-wind to be ejected from an inner radius $R_{\rm in}$ to an 
outer radius $R_{\rm ex}$.
In this case the stellar magnetic field lines are connected to the disk
only in a small region around $R_{\rm in}$ (see Fig. \ref{fig1}).
$R_{\rm in}$ can be determined as the distance 
from the star where the magnetic stress $B_{\phi} B_p R_0^2$ begin to dominates over the 
viscous stress $\dot{M} d\left(\Omega R_0^2\right)/dR_0$ \citep[e.g.][]{wan96}:
\begin{equation}
  R_{\rm in} = B_p(t)^{4/7} \left[
          \frac{2 R_\star^6}
               {\dot{M}\left(GM_\star\right)^{1/2}} \right]^{2/7}
  \label{eq:rin}
\end{equation}
where
$R_\star$ and $M_\star$ are the star radius and mass,
$\Omega=(GM/R_0^3)^{1/2}$, and we assume $B_{\phi} \approx B_p$ 
(with $B_p \propto m$, defined by eq. \ref{eq:bvar}) and a 
full penetration of the vertical magnetic field lines into the disk.
The same dependence of $R_{\rm in}$ on $B_p$ is 
recovered and the results obtained in the previous section are unchanged.

Alternatively the stellar-disk magnetic field lines
might close again via magnetic reconnection
producing quasi-periodic ejections
\citep[e.g.][]{goo99,mat02}.
Our results correspond to the hypothesis
that the mass flux carried by the quasi-periodic ejections is much less
then the mass flux carried by the disk-wind.
If on the other side the mass flux carried by the coronal mass ejection is 
important, the stellar cycle would also produce a variation in the jet velocity.
Actually, in contrast with the disk-wind case, for a
coronal mass ejection an increase in the magnetic field would produce 
an increase in the ejection velocity (as found in simulations
by \citealt{mat02}). In fact, in coronal mass ejections the magnetic energy is 
directly transformed in kinetic energy by reconnection events.

We concentrated on a cyclic magnetic field as
origin of the variations in the inner launching radius for the disk
wind. However, any physical process that leads to quasi-periodic
variations of the launching region will result in similar variations
of the mean asymptotic velocity -- and therefore to the appearance
of knots -- as long as those variations are large enough.

It is unclear which process (other than a stellar magnetic field
variation) might create variations on a 
$\sim 1-20$ yr timescale.
\citet{st03} showed that a relaxation oscillator 
may produce a change in the inner radius in X-ray binaries. The timescale
of the oscillation depends on the details of the stellar magnetosphere
and can vary by two orders of magnitude.
A variation in the accretion rate may also produce a similar
variation in the inner radius (see eq. \ref{eq:rin}, and the discussion
by \citealt{bla01}).

\citet{har04} showed that short period T Tauri binaries do not seem 
to have jet properties different from those of wider binaries or single
stars, arguing in this way against binarity as the origin of jet knots.
\citet{har04} also argued that there is no 
correlation between the
ejection of a new knot in CW Tau and an increase in source brightness.
One might have expected this if the knot was caused by an accretion outburst.
Finally, \citet{fer06} suggested that the alternation between X- and 
Y-type interactions between stellar and disk magnetic fields 
as due to a magnetic field cycle could explain the observed periodicity.


\section{Conclusions}
\label{conclusion}

In this paper we explored the effect of a stellar magnetic field 
variation on the ejection velocity of protostellar jets.
While large scale knots may successfully be explained by a large
increase in the accretion rate, we showed that small scale knots 
may be produced by changes in the inner radius.

We showed that a stellar magnetic field variation,
if present, represents a natural candidate to produce stellar 
knots similar to the observed.
In fact, a stellar magnetic field variation produces a change
in the inner radius and therefore in the jet velocity.
We also estimated that a stellar magnetic field variation 
$\lesssim$ 50\% produces a variation ($\sim$ 20\%) in 
the average velocity large enough to produce knots similar
to the observed.

The stellar magnetic field periodic or quasi-periodic variation
remains an hypothesis of our scenario,
but future large scale temporal observations of magnetic
field in young stars, and progress in dynamo theory, 
will both help to provide an answer to this problem.

\acknowledgments
   We thank Turlough Downes, Jonathan Ferreira, Tom Ray and Nektarios Vlahakis for useful discussions.
   We also acknowledge Edith Salado Lopez for the elaboration of Fig. \ref{fig1}.
   Part of this work was supported by the European Community's Marie 
   Curie Actions - Human Resource and Mobility within the JETSET (Jet 
   Simulations, Experiments and Theory) network under contract MRTN-CT-2004 
   005592, and by the Agence Nationale de la Recherche (ANR).

\end{document}